\documentclass[aps,pra,floatfix,twocolumn,showpacs,english,10pt]{revtex4-1}
\pdfoutput=1

\usepackage{babel}                   
\usepackage[utf8]{inputenc}          %
\usepackage[T1]{fontenc}             %
\usepackage{lmodern}                 
\usepackage{graphicx}                
\usepackage{amsmath, mathtools}      
\usepackage{amsfonts, amssymb}       %
\usepackage{hyperref}                
\usepackage{cleveref}                

\graphicspath{{images/}}

\newcommand{\ket}[1]{\left| #1 \right\rangle}

\newcommand{\braopket}[3]{\left\langle #1 \middle| #2 \middle| #3 \right\rangle}
\newcommand{\parens}[1]{{\left( #1 \right)}}

\newcommand{\braces}[1]{{\left\lbrace #1 \right\rbrace}}
\newcommand{\opn}[1]{\operatorname{#1}}
\newcommand{\qtext}[1]{\quad\text{#1}\quad}
\newcommand{\sgn}{{\opn{sgn}}}
\newcommand{\unity}{{\mathbb{I}}}
\newcommand{\dunity}{{\mathbb{J}}}
\newcommand{\dft}{{\opn{DFT}}}
\newcommand{\dfto}{{\widehat{\opn{DFT}}}}
\newcommand{\dsti}{{\opn{DST-I}}}
\newcommand{\dstiii}{{\opn{DST-III}}}
\newcommand{\dstiiip}{{\opn{\overline{DST}-III}}}
\newcommand{\dstio}{{\opn{\widehat{DST}-I}}}
\newcommand{\dstiiio}{{\opn{\widehat{DST}-III}}}
\newcommand{\dddots}{{\mathpalette\ddddots\relax}}
\newcommand{\ddddots}{\reflectbox{$\ddots$}}
\newcommand{\ci}{\mathcal{C}^{\mathrm{I}}}
\newcommand{\ciii}{\mathcal{C}^{\mathrm{III}}}

\begin{document}

\title{Implementing the sine transform of fermionic modes as a tensor network}

\author{Hannes Epple} \email{hannes.epple@stud-mail.uni-wuerzburg.de}
\author{Pascal Fries}
\author{Haye Hinrichsen}
\affiliation{
  Fakult\"at f\"ur Physik und Astronomie, \\
  Julius-Maximilians Universit\"at W\"urzburg, \\
  Am Hubland, 97074 W\"urzburg, Germany
}%

\date{\today}

\begin{abstract}
  Based on the algebraic theory of signal processing, we recursively decompose the discrete sine
  transform of first kind {($\dsti$)} into small orthogonal block operations.  Using a diagrammatic
  language, we then second-quantize this decomposition to construct a tensor network implementing
  the {$\dsti$} for fermionic modes on a lattice.  The complexity of the resulting network is shown to
  scale as $\frac 54 n \log n$ (not considering swap gates), where $n$ is the number of lattice
  sites.  Our method provides a systematic approach of generalizing Ferris' spectral tensor network
  for non-trivial boundary conditions.
\end{abstract}


\maketitle

\section{Introduction}
The study of tensor networks is currently a topic of growing interest both in condensed matter
physics and quantum computation.  In condensed matter physics, tensor networks can be used to model
the entanglement structure of quantum states and are therefore well suited for the study of ground
states of strongly correlated systems \cite{Eisert:13,Orus:14a,Orus:14b}.  Specifically, the
formulation of the \emph{multiscale entanglement renormalization ansatz} (MERA) in this framework
has shown to be very fruitful \cite{Vidal:08,Evenbly:09,Carboz:09,Orus:14b} and can be understood as
a lattice realization of the \emph{holographic principle} \cite{Susskind:95,Swingle:12},
contributing to a better understanding of gauge-gravity type dualities \cite{Maldacena:98}.  In
quantum computation, on the other hand, tensor networks are known as \emph{quantum circuits}
\cite{Nielsen:00} and provide a natural framework for the decomposition of a large, complicated
unitary operation into a sequence of small local building blocks -- a simple example being the
factorization of the \emph{quantum Fourier transform} \cite{Nielsen:00} into a sequence of Hadamard
and phase gates, which causes the efficiency of Shor's algorithm \cite{Shor:97}.

A unification of both subjects---namely, the simulation of strongly correlated quantum systems by a
quantum circuit---was already proposed by Feynman in 1982 \cite{Feynman:82} and has recently been
realized by a circuit \cite{Verstraete:09} for implementing the \emph{exact} dynamics of free
fermions on a circle.  More recently, Ferris refined this idea \cite{Ferris:14}, giving it the
interpretation of a \emph{spectral tensor network}, which implements the Fourier transform of
\emph{fermionic modes}, hence diagonalizes the Hamiltonian of free fermions
\cite{Lieb:61}.  Additionally, the geometry of this network generalizes to higher dimensions, while
always staying easily contractible, such that it can be used for the \emph{classical simulation} of
phase transitions in more than one dimension -- a feature that is absent in the MERA
\cite{Evenbly:14a,Evenbly:14b}.

The starting point for the construction of Ferris' spectral tensor network is a recursive algorithm
for the discrete Fourier transformation (DFT), widely known as Fast Fourier transform (FFT)
\cite{Cooley:65}.  The network is then chosen such that it implements the FFT in the one-particle
sector of the theory \cite{Ferris:14}.  This means that \emph{every particle} is transformed
separately, subject to the constraint of their respective indistinguishability.  Note that this is
entirely different from the quantum Fourier transform, which calculates a single DFT on the space
spanned by all particles.

A closely related transformation, on which we will focus in the present work, is the discrete sine
transformation of first kind {($\dsti$)}.  This is a variant of the DFT for vanishing Dirichlet boundary
conditions and thus indispensable for diagonalization procedures of systems on an open chain
\cite{Lieb:61}.  However, the boundary conditions break the convenient cyclic
translational symmetry vital to the decomposition of the DFT, meaning that the original FFT
algorithm can no longer be used.

The \emph{algebraic theory of signal processing} \cite{Moura:08a,Moura:08b} tackles this problem by
constructing an algebraic framework for spectral transformations.  Based on these notions, decompositions
of whole classes of transformations were derived in \cite{Moura:03,Moura:08c}, including the {$\dsti$}
as a special case.  However, in contrast to the simple case of the standard FFT, the resulting
decomposition almost entirely consists of non-unitary elementary transformations.  While this is
acceptable for numerical recipes restricted to a single particle, we run into problems when
translating the decomposition into a quantum circuit, implementing the transformation for
indistinguishable particles.

In this paper we unitarize a recursive algorithm for the {$\dsti$}, originally given in \cite{Moura:08c},
and show how to second-quantize the resulting algorithm, obtaining a spectral tensor network for a
fermionic chain with open ends.  To this end we reorganize the network in a non-trivial way.  To keep
the formal ballast as small as possible, the present paper explains most of the steps in terms of
diagrams, showing selected parts of the decomposition and an explicit example of a complete algorithm.

\section{The fast Fourier transform and diagram notation}
The conventional discrete Fourier transform (DFT) converts a sequence of $n$ complex numbers
$x_0, x_1, \ldots, x_{n-1}$ into another sequence of complex numbers
$\tilde x_0, \tilde x_1 ,\ldots, \tilde x_{n-1}$ by means of the linear transformation
\[
  \tilde x_a := \sum_{b=0}^{n-1} \mathrm{e}^{-2 \pi \mathrm{i} a b / n} x_b.
\]
Defining the phase factor $\omega_n := \mathrm{e}^{-2\pi \mathrm{i} / n}$, this transformation can
be written as
\[
  \tilde x = \dft_n x
  \qtext{with}
  \dft_n^{ab} = \omega_n^{a \cdot b}.
\]
Here and in the following, we use the convention that the lower index $n$ denotes the dimension of
the respective vector space, while upper indices $0 \leq a,b < n$ label components of the matrix.

For even $n=2m$, the recursion formula
\begin{subequations}
\begin{multline} \label{eq:fft-formula}
  \dft_{2m} = L_{2m} (\dft_m \oplus \dft_m) \\
  \times \parens{\unity_m \oplus \opn{diag}(\omega_{2m}^0,\ldots, \omega_{2m}^{m-1})}
  (\dft_2 \otimes \unity_m)
\end{multline}
with
\begin{gather*}
  \unity_m^{ab} := \delta_{ab}, \quad 
  L_{2m} := \bar{L}_{2m-1} \oplus 1, \quad \text{and} \\
  \bar{L}_{2m-1}^{ab} :=
  \begin{cases}
    1 &\text{iff} \quad b \equiv am \mod (2m - 1) \\
    0 &\text{otherwise}
  \end{cases} 
\end{gather*}
is known as the radix 2 fast Fourier transform (FFT) \cite{Cooley:65}.  It factorizes the $\dft_{2m}$
into three pieces: The rightmost factor $\dft_2 \otimes \unity_m$ is a basis transformation,
consisting of $m$ copies of the matrix
\[
  F := \dft_2 = \begin{pmatrix} 1 & 1 \\ 1 & -1 \end{pmatrix}
\]
acting on the pairs of components $(\ell, m+\ell)$, $\ell=0,\ldots,m-1$.  The next factor,
\[
  \dft_m \oplus \dft_m \opn{diag}(\omega_{2m}^0,\ldots, \omega_{2m}^{m-1}),
\]
is a direct sum of two $\dft_m$, one of which is modified by multiplication with a diagonal
matrix of so called \emph{twiddle factors}.  The last factor $L_{2m}$ just permutes the basis
vectors, sorting them into odd and even portions.

Diagrammatically, the recursion relation \eqref{eq:fft-formula} for $n=8$ can be represented as
\begin{equation} \label{eq:fft-diagram}
  \dft_8 \; =
  \begin{gathered}
    \includegraphics[scale=1.0]{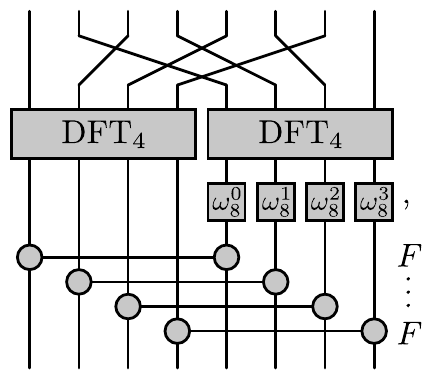}
  \end{gathered}
\end{equation}
\end{subequations}
running from bottom to top.  Here, blocks represent matrices, the ingoing lines are columns, while
the outgoing lines lines are rows.  The composition rules for such diagrams are
\begin{equation}
  \label{eq:single-particle-diagram-rules}
  \begin{gathered}
    \includegraphics[scale=1.0]{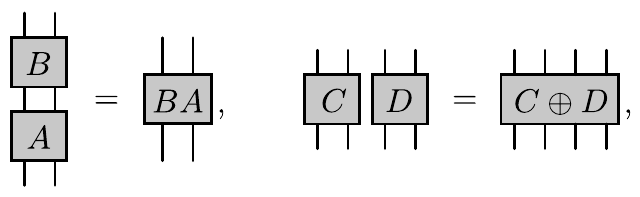}
  \end{gathered}
\end{equation}
i.e., vertical composition couples rows of the lower block to columns of the upper block and
therefore represents ordinary matrix multiplication, while horizontal composition gives rise to the
direct sum of matrices \footnote{Note that this is in strict contrast to the usual diagrammatics of
  tensor networks and stems from the fact that we are not considering multiple particles yet. This
  will change in \cref{sec:diagram-quantization}, where horizontal composition will give rise to the
  tensor product. We thank the referee for suggesting to emphasize this point.}.  Since the matrix
$F$ is applied to non-neighboring lines, we do not draw it as a box but rather use the shorthand
notation
\begin{equation}
  \label{eq:shorthand-hline}
  \begin{gathered}
    \includegraphics[scale=1.0]{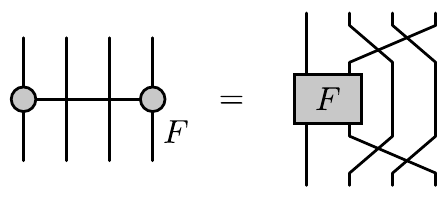}
  \end{gathered}
\end{equation}
with two bullets to indicate on which lines the matrix acts.  Note that lines can be crossed
arbitrarily, allowing us to move the remaining unaffected lines freely.

Now, if $n$ is a power of $2$, we can use \cref{eq:fft-formula}, anchored at $\dft_2 = F$, to
implement the $\dft_n$ using only $2 \times 2$-matrices.  In the above diagrammatic language, this
amounts to having no blocks act on more than two lines.  By construction, the number of blocks in
such a diagram then scales as $n \log_2 n$, which therefore serves as an upper bound for the
computational complexity of the DFT.

In order to interpret the DFT as a change of orthonormal bases in a single-particle Hilbert space,
it needs to be unitary.  This is achieved by the normalization
\[
  \dfto_n := \frac1{\sqrt n} \dft_n
\]
and, as can be easily checked, \cref{eq:fft-formula,eq:fft-diagram} remain valid under the
replacement $\dft \to \dfto$, provided we also normalize $F$ by
\begin{equation}
  \label{eq:def-f-hat}
  F \to \hat{F} := \frac 1{\sqrt 2} F.
\end{equation}
We thus obtain a decomposition of the large unitary $\smash{\dfto_n}$ into small unitary building
blocks, which again carry the interpretation of basis transformations in the single particle Hilbert
space.

The process of second quantization \cite{Berezin:66} then maps the $\smash{\dfto_n}$ to a basis
change in the many-particle Hilbert space. Remarkably, the FFT scheme \eqref{eq:fft-diagram} is
still valid in the many-particle case, provided that we slightly change the composition rules
\eqref{eq:single-particle-diagram-rules} and replace the blocks by their respective second
quantizations. We will discuss second quantization of Fourier transforms in more detail later in
\cref{sec:diagram-quantization}.

\section{Decomposing the discrete sine transform}
The discrete sine transform (DST) is a real linear transformation which captures essentially the
imaginary part of the DFT.  As it expands the data in sinusoids, this transformation is particularly
useful for discrete systems with Dirichlet boundary conditions.  However, depending on the
implementation of the boundary conditions and the respective periodic continuation, one
distinguishes various types of DSTs, which are usually labeled by Roman numbers from I to VIII
\cite{Moura:08b}.

We focus on the {$\dsti$} here, which corresponds to a periodic continuation that is odd around both
$x_{-1}$ and $x_n$.  Thus, we have $x_{-1}=x_n=0$, making the {$\dsti$} suitable for systems with open
boundary conditions.  As we will see below, for a suitable recursion scheme we also need the
{$\dstiii$}.  The two DSTs are defined by the matrices
\begin{align}
  \label{eq:dst-1-matrix}
  \dsti_{n}^{ab} &=  \sin \frac{(a+1)(b+1)\pi}{n+1} \quad \text{and} \\
  \label{eq:dst-3-matrix}
  \dstiii_{n}^{ab} &=  \sin \frac{(a+\frac{1}{2})(b+1)\pi}{n},
\end{align}
with $0 \leq a,b < n$, which are non-orthogonal.  An advantage of considering the {$\dsti$} is that
we only need the {$\dstiii$} and the {$\dsti$} itself in the corresponding recursion.  In principle,
it is possible to consider other types of sine (or cosine) transforms and orthogonalize them in a
similar way as described in the next section, though the corresponding recursions may be more
complicated.  Specifically, the {$\dsti$} of odd size $n=2m-1$ can be expressed recursively in terms
of {$\dsti$} and {$\dstiii$} as \cite{Moura:08c}
\begin{subequations}
\begin{multline}
  \label{eq:dst-1-alg}
  \dsti_{2m-1} = \\ 
  \bar{L}_{2m-1} \big( \dstiii_m \oplus \dsti_{m-1} \big) B_{2m-1}.
\end{multline}
Here, the rightmost factor is a base change matrix
\begin{equation*}
  B_{2m-1} :=
  \begin{pmatrix}
    \unity_{m-1} & & \phantom{-}\dunity_{m-1} \\
     & 1 & \\
    \unity_{m-1} & & -\dunity_{m-1} 
  \end{pmatrix}
\end{equation*}
with the $(m-1)\times(m-1)$ identity matrix $\unity_{m-1}$ and
\begin{equation*}
  \dunity_{m-1} :=
  \begin{pmatrix}
     & & 1 \\
     & \dddots & \\
    1 & &
  \end{pmatrix}
  .
\end{equation*}
Note that $B_{2m-1}$ can be split into an interaction part and a permutation,
\[
  B_{2m-1} = 
  \begin{pmatrix}
    \unity_{m-1} & & \phantom{-}\unity_{m-1} \\
     & 1 & \\
    \unity_{m-1} & & -\unity_{m-1} 
  \end{pmatrix}
  \cdot
  \begin{pmatrix}
    \unity_{m-1} & & \\
     & 1 & \\
     & & \dunity_{m-1} 
  \end{pmatrix},
\]
where the interaction part acts on pairs of components $(\ell,m+\ell)$, $\ell=0,\dots,m-2$ via $F$.
The middle factor is just a direct sum of a {$\dstiii$} and a {$\dsti$} of smaller sizes $m$ and
$m-1$, respectively, while the leftmost factor is a permutation defined in the previous section.

In the diagrammatic notation established before, the recursion relation \eqref{eq:dst-1-alg} for
$n=7$ can be represented as
\begin{equation}
  \label{eq:diag-dst-1-alg}
  \dsti_7 = 
  \begin{gathered}
    \includegraphics[scale=1.0]{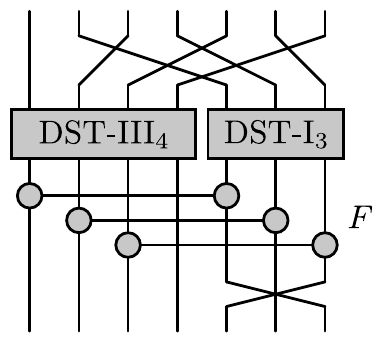}
  \end{gathered}.
\end{equation}
\end{subequations}

The {$\dstiii$} appearing in this recursion can be further reduced by means of another
recursion relation.  For the {$\dstiii$} of size $n=2m$ we use 
\begin{subequations}
\begin{multline}
  \label{eq:dst-3-alg}
  \dstiii_{2m} = K_{2m} \big( \dstiii_m \oplus \dstiii_m \big) \\ 
  \times \big( X_m^- \oplus X_m^+ \big) \big( \dstiiip_2 \otimes \, \unity_m \big) \bar{B}_{2m},
\end{multline}
which is a special case of a more general decomposition for $n=qm$ \cite{Moura:08c}.  Note that in
contrast to the binary recursion \eqref{eq:dst-1-alg}, the above formula recurs to two copies of
{$\dstiii$} itself.  Again, the rightmost factor is a base change matrix
\begin{equation*}
  \bar{B}_{2m} :=
  \begin{pmatrix}
    \unity_{m-1} & & \dunity_{m-1} \\
     & 1 & \\
     & & \unity_{m-1}
  \end{pmatrix}
  \oplus 1
  ,
\end{equation*}
which leaves the components $m-1$ and $2m-1$ unaffected and applies the matrix  
\begin{equation*}
  A :=
  \begin{pmatrix}
    1 & 1 \\
    0 & 1
  \end{pmatrix}
\end{equation*}
to pairs of components $(\ell,2m-2-\ell)$, $\ell=0,\dots,m-2$.  Because of its tensor product
structure, the next factor $\dstiiip_2 \otimes \, \unity_m$ applies a scaled {$\dstiii$} to pairs of
components $(\ell,m+\ell)$, $\ell=0,\dots,m-1$, given by the matrix
\[
  C := \dstiiip_2 = F \opn{diag}(1,\sqrt{2}).
\]
The third factor in \cref{eq:dst-3-alg} is a direct sum of two matrices $X_m^\pm$, which will be
discussed later on.  The next factor consists of a direct sum of two smaller {$\dstiii$s}, while the
leftmost factor is again a permutation, defined by
\[
  K_{2m} := ( \unity_2 \oplus \dunity_2 \oplus \unity_2 \oplus \dunity_2 \oplus \dots ) L_{2m}.
\]
We can represent the recursion relation \eqref{eq:dst-3-alg} diagrammatically as
\begin{equation}
  \label{eq:diag-dst-3-alg}
  \dstiii_8 =
  \begin{gathered}
    \includegraphics[scale=1.0]{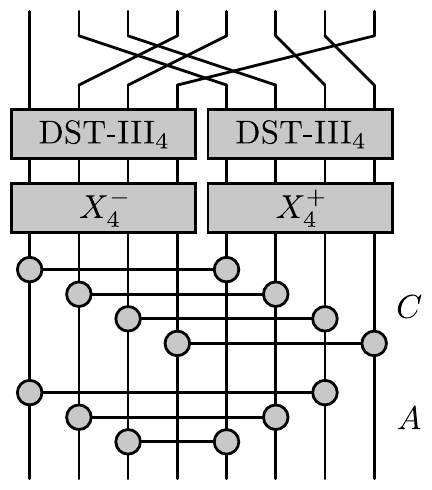}
  \end{gathered}.
\end{equation}
\end{subequations}

The remaining parts to consider are the matrices $X_m^\pm$.  For even size $m$, they are given by
\begin{equation}
  \label{eq:x-matrix}
  X_m^\pm :=
  \begin{pmatrix}
    c_m^1 & & & & \hspace{-.2cm} s_m^{\pm (m-1)} & 0 \\
     & \hspace{-.2cm} \ddots & & \hspace{-.4cm} \dddots & & \\
     & & \hspace{-.2cm} c_m^{m/2}+s_m^{\pm m/2} & & & \vdots \\
     & \hspace{-.2cm} \dddots & & \hspace{-.4cm} \ddots & & \\
    s_m^{\pm1} & & & & \hspace{-.2cm} c_m^{m-1} & 0 \\
    0 & & \hspace{-.3cm} \cdots & & 0 & c_m^m \\
  \end{pmatrix}
\end{equation}
with
\[
  c_m^\ell := \cos \frac{\ell\pi}{4m} \qtext{and}
  s_m^\ell := \sin \frac{\ell\pi}{4m}.
\]
Clearly, these matrices can also be decomposed into blocks acting only on pairs of components
$(\ell-1,m-1-\ell)$, $\ell=1,\dots,m/2-1$ via
\begin{equation}
  \label{eq:y-matrix}
  Y^\pm_{\ell,m} :=
  \begin{pmatrix}
    c_m^\ell & s_m^{\pm (m-\ell)} \\
    s_m^{\pm \ell} & c_m^{m-\ell}
  \end{pmatrix}
  .
\end{equation}
Further, in the center and the right lower corner of the matrix \eqref{eq:x-matrix}, we have the
factors
\[
  y^\pm := c_m^{m/2}+s_m^{\pm m/2} = \sqrt{1 \pm \frac{1}{\sqrt{2}}}
  \qtext{and} c_m^m = \frac{1}{\sqrt{2}},
\]
acting on the components $m/2-1$ and $m-1$.  For example, the matrix $X_m^\pm$ of size $m=8$ can be
drawn in our diagrammatic notation as
\begin{equation*}
  X_8^\pm = 
  \begin{gathered}
    \includegraphics[scale=1.0]{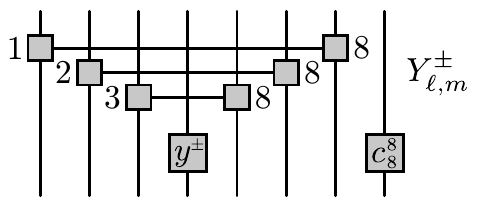}
  \end{gathered}
  ,
\end{equation*}
where the indices $\ell$ and $m$ of the $2 \times 2$ matrices $Y_{\ell,m}^\pm$ are given on the left
and right of the corresponding symbol, respectively.

Putting all this together, the two mutually dependent recursions relations \eqref{eq:dst-1-alg} and
\eqref{eq:dst-3-alg} together with the closing conditions
\[
  \dsti_1 = 1 \qtext{and}
  \dstiii_2 = F \opn{diag} \Big( \frac{1}{\sqrt{2}},1 \Big)
\]
allow us to calculate the {$\dsti$} of size $n=2^k-1$ using only $2 \times 2$ matrices.  However, in
the existing formulation, all the occurring matrices, except for permutations, are still
non-orthogonal.  This is a problem in the corresponding quantum version, since non-orthogonal
building blocks cannot be interpreted as elementary changes of orthonormal bases.  Fortunately, it
is possible to reformulate the recursion relations in an orthogonal way, as will be shown in the
next section.

\section{Orthogonalization of the recursion relations}
\begin{figure*}
  \centering
  \includegraphics[scale=1.0]{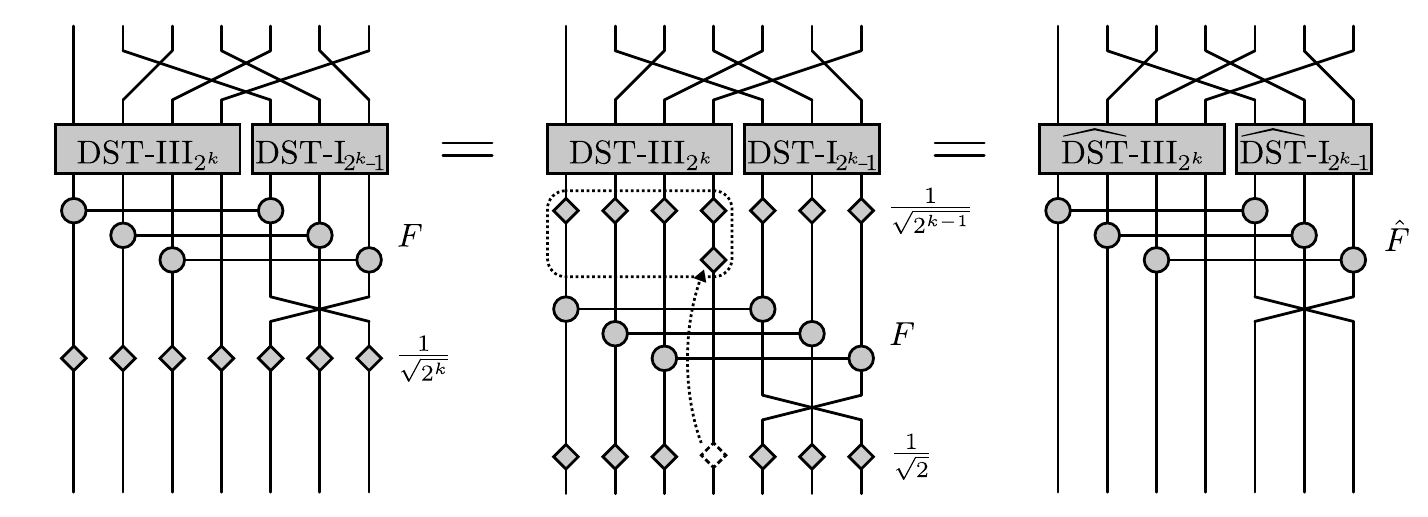}
  \caption{Orthogonalization of the recursion relation \cref{eq:dst-1-alg} for the {$\dsti$} of size
    $n=2^{k+1}-1$, shown here for $k=2$.}
  \label{fig:dst-1-orthogonal}
\end{figure*}

We now show step by step how to obtain an orthogonal version of the recursion relations
\eqref{eq:dst-1-alg} and \eqref{eq:dst-3-alg} discussed in the previous section.  For convenience,
we label all orthogonal matrices by a hat symbol.

As stated before, the DSTs defined by \cref{eq:dst-1-matrix,eq:dst-3-matrix} are non-orthogonal.
However, all DSTs can be made orthogonal by a suitable scaling of rows and columns.  Let us start
with the {$\dsti$}.  Multiplying the corresponding matrix \eqref{eq:dst-1-matrix} by its transpose, we
find that the correct scaling is given by
\begin{subequations} \label{eq:dst-1-ortho}
  \begin{equation}
    \dstio_n := \dsti_n \sqrt{\frac{2}{n+1}},
  \end{equation}
  meaning that all matrix entries are rescaled identically.  Representing multiplication of a
  component by a small diamond this relation may be drawn as
  \begin{equation}
    \begin{gathered}
      \includegraphics[scale=1.0]{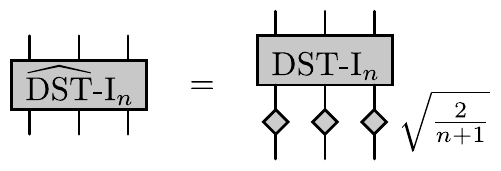}
    \end{gathered}
  \end{equation}
\end{subequations}
in our graphical notation.  Now we can try to recast the recursion \eqref{eq:dst-1-alg} in terms of
orthogonal matrices.  The procedure is shown in \cref{fig:dst-1-orthogonal} for a {$\dsti$} of size
${n=2^{k+1}-1}$.

The diagram on the left is just \cref{eq:diag-dst-1-alg} with the proper scaling factors according
to \cref{eq:dst-1-ortho} added at the bottom.  In the first step, we split up the factors for each
component and move factors of $1/\sqrt{2^{k-1}}$ above the matrices~$F$.  This is possible since we
have the same factor at every component.  As no matrix $F$ is acting on the component in the middle,
we can also move the remaining factor of $1/\sqrt{2}$ for this component further up, as indicated by
the arrow.  Now, all we have to do is to absorb all factors into the matrices directly above them.
We obtain orthogonal matrices $\hat{F}$ as defined in \cref{eq:def-f-hat}.  Further, the factors
absorbed into the {$\dsti$} are just the ones from \cref{eq:dst-1-ortho}, so we obtain an orthogonal
version of this transform.  Since we know that the whole transform in \cref{fig:dst-1-orthogonal} is
orthogonal and all other building blocks are orthogonal, too, we can conclude that the {$\dstiii$}
together with the factors in the dotted box must also be orthogonal.  Defining the orthogonalized
{$\dstiii$} as
\begin{subequations}
  \label{eq:dst-3-ortho}
  \begin{equation}
    \dstiiio_n :=
    \dstiii_n \sqrt{\frac 2n} \opn{diag} \Big( 1,\dots,1,\frac{1}{\sqrt{2}} \Big) 
  \end{equation}
  or diagrammatically as
  \begin{equation}
    \begin{gathered}
      \includegraphics[scale=1.0]{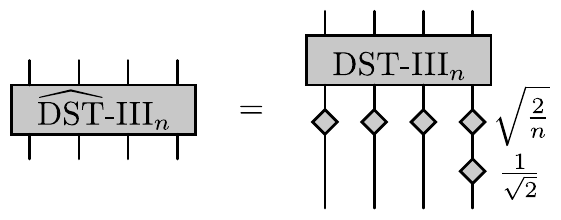}
    \end{gathered},
  \end{equation}
\end{subequations}
we obtain the right diagram of \cref{fig:dst-1-orthogonal}, where all occurring matrices are
orthogonal.

\begin{figure*}
  \centering
  \includegraphics[scale=1.0]{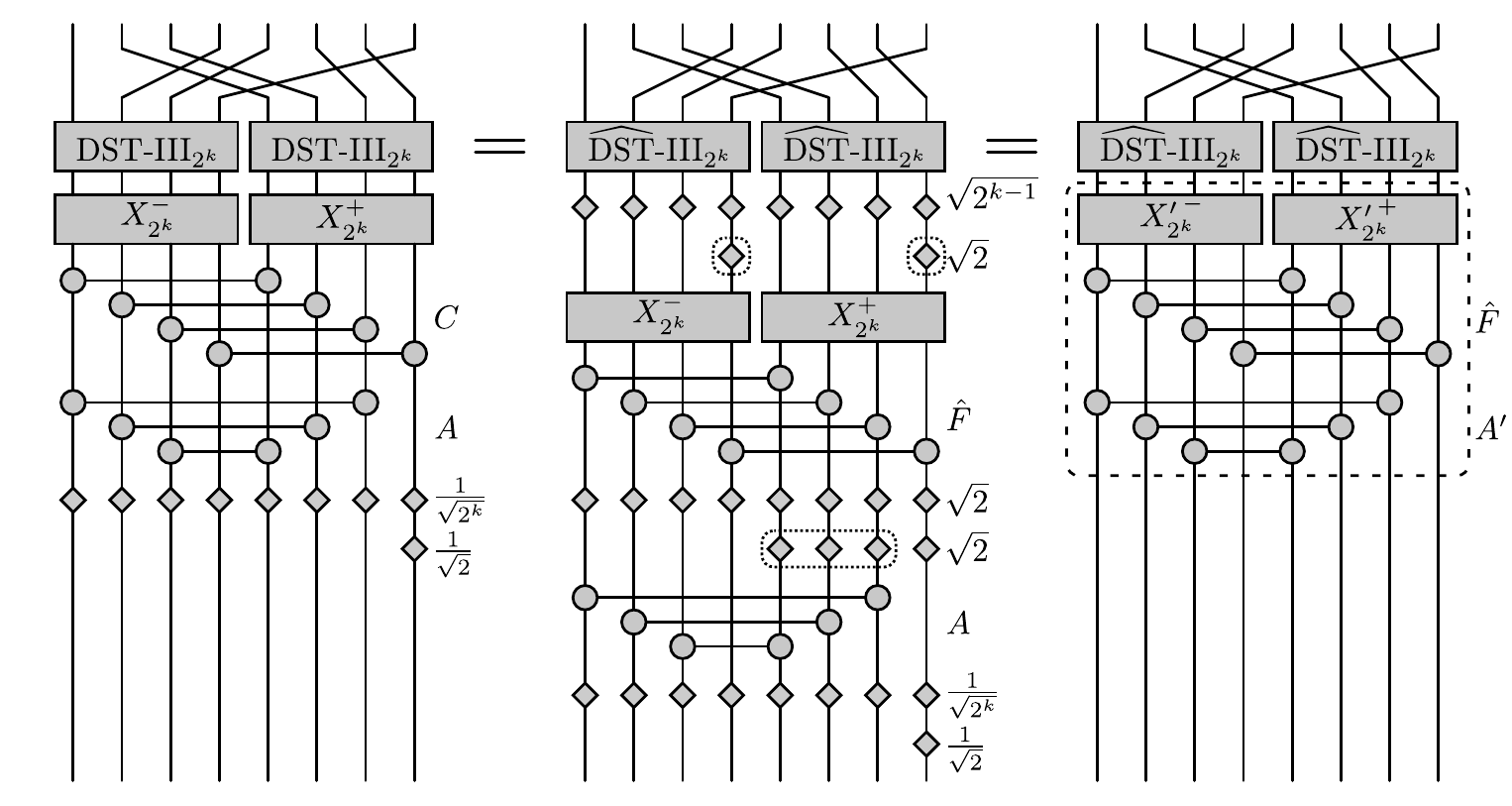}
  \caption{Orthogonalization of the recursion relation \eqref{eq:dst-3-alg} for the {$\dstiii$} of
    size $n=2^{k+1}$.  Again, we have $k=2$.  The part in the dashed box in the right diagram is
    considered in \cref{fig:dst-3-orthogonal-detail}.}
  \label{fig:dst-3-orthogonal}
\end{figure*}

Knowing the orthogonal version of the {$\dstiii$}, we now turn to the corresponding recursion
relation \eqref{eq:dst-3-alg}.  In a first step, we replace the {$\dstiii$} by its orthogonal
version, as it is shown in \cref{fig:dst-3-orthogonal} for a {$\dstiii$} of size $n=2^{k+1}$.  We
start with the recursion relation from \cref{eq:diag-dst-1-alg} with scaling factors according to
\cref{eq:dst-3-ortho} added at the bottom of the diagram.  In the first step, we use
\cref{eq:dst-3-ortho} to replace the two smaller DSTs by their orthogonal versions and the
corresponding inverse scaling factors.  Further, we express $C = F \opn{diag} (1, \sqrt 2)$ in terms
of $\hat{F}$ using \cref{eq:def-f-hat}.  All factors except the ones in the dotted boxes cancel out.
In the second step, we absorb those into the matrices below, obtaining the matrices
\begin{gather}
  {X'}_{\!\!m}^{\pm} := \opn{diag} (1,\dots,1,\sqrt{2})\, X_m^\pm \label{eq:x'-def}\\
  \text{and} \quad A' := \opn{diag} (1,\sqrt{2})\, A, \nonumber
\end{gather}
which unfortunately are still non-orthogonal.  The part of the recursion relation that remains to be
orthogonalized is indicated by a dashed box in the right diagram of \cref{fig:dst-3-orthogonal}.

\begin{figure*}
  \centering
  \includegraphics[scale=1.0]{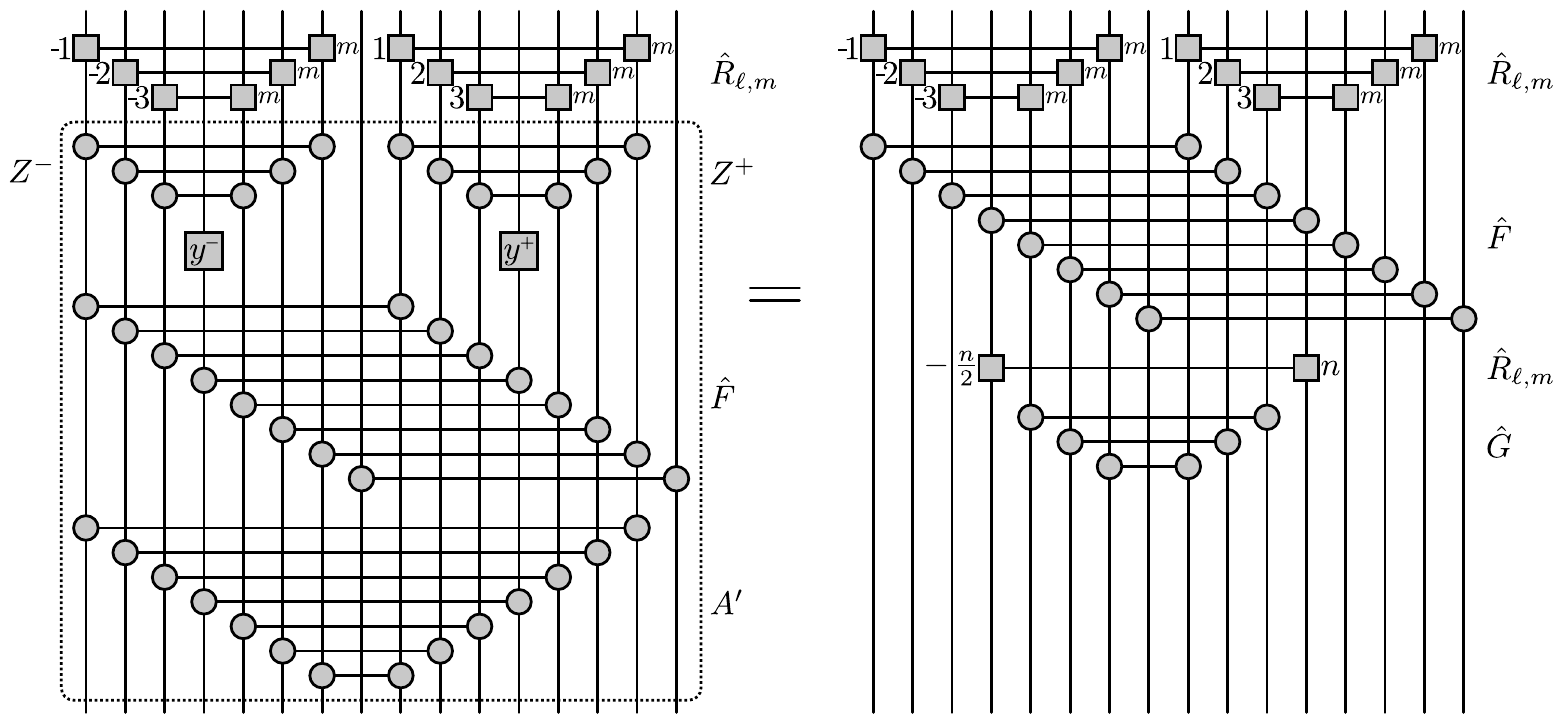}
  \caption{Detail for the orthogonalization of the recursion relation \eqref{eq:dst-3-alg} for the
    $\dstiii$ of size $n=2^{k+1}$, shown here for $k=3$, in order to resolve all relevant details.
    This figure corresponds to the part in the dashed box in the right diagram of
    \cref{fig:dst-3-orthogonal} with $m=2^k$.  The dotted box in the diagram on the left hand side
    of this figure indicates the part that remains to be reformulated in terms of orthogonal
    operations, as shown on the right hand side.}
  \label{fig:dst-3-orthogonal-detail}
\end{figure*}

Let us have a look at ${X'}_{\!\!m}^\pm$ first.  The diagonal matrix in \cref{eq:x'-def} just cancels
the factor $c_m^m=1/\sqrt{2}$ in the lower right corner of $X_m^\pm$.  Further, the occurring
matrices $Y_{\ell,m}^\pm$, defined in \cref{eq:y-matrix}, can be decomposed as
\[
  Y_{\ell,m}^\pm = \hat{R}_{\pm\ell,m} Z^\pm,
\]
with the rotation matrix
\[
  \hat{R}_{\ell,m} := 
  \begin{pmatrix}
    \cos \frac{\ell\pi}{4m} & -\sin \frac{\ell\pi}{4m} \\
    \sin \frac{\ell\pi}{4m} & \phantom{-}\cos \frac{\ell\pi}{4m}
  \end{pmatrix}
\]
and the non-orthogonal matrix
\[
  Z^\pm := 
  \begin{pmatrix}
    1 & \pm1/\sqrt{2} \\
    0 & \phantom{\pm}1/\sqrt{2}
  \end{pmatrix}.
\]
Thus, the diagrammatic representation of ${X'}_{\!\!m}^{\pm}$, drawn below for $m=8$, is
\begin{equation*}
  {X'}_{\!\!8}^\pm = 
  \begin{gathered}
    \includegraphics[scale=1.0]{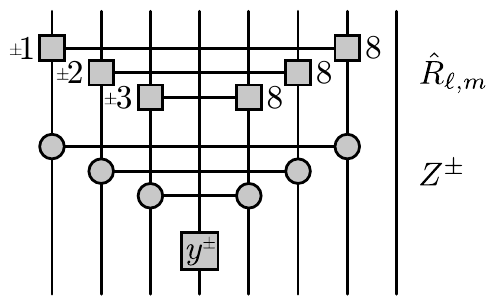}
  \end{gathered}.
\end{equation*}
Using this representation, the operations in the dashed box in the right diagram of
\cref{fig:dst-3-orthogonal} can be redrawn as shown in the left diagram of
\cref{fig:dst-3-orthogonal-detail}.  We have doubled the number of components in this diagram to
show all relevant structures and used the abbreviation $m=2^k$.  Again, the non-orthogonal part is
highlighted by a dotted box.

In order to obtain the expression made up from orthogonal matrices given by the right diagram in
\cref{fig:dst-3-orthogonal-detail}, we observe that the part in the dotted box can be decomposed
into three sorts of blocks for any size $n=2^{k+1}$.  On the pair of components $(2^k-1,2^{k+1}-1)$,
we have a trivial block
\begin{equation*}
  \begin{gathered}
    \includegraphics[scale=1.0]{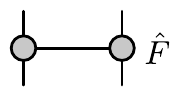}
  \end{gathered},
\end{equation*}
which is already orthogonal.  Further, the pair of components
$(2^{k-1}-1,2^k+2^{k-1}-1)$ is coupled by
\begin{equation*}
  \begin{gathered}
    \includegraphics[scale=1.0]{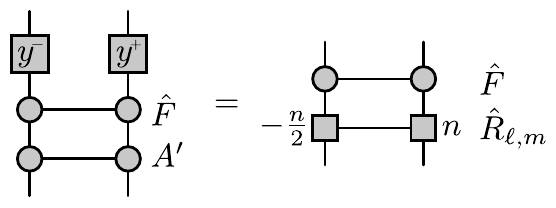}
  \end{gathered},
\end{equation*}
where the reformulation on the right hand side only contains matrices which are orthogonal.  All
other operations decompose into blocks acting on four components
\[
  (\ell,2^k-2-\ell,2^k+\ell,2^{k+1}-2-\ell)
\]
with $\ell=0,\dots,2^{k-1}-2$.  These blocks may be orthogonalized by the relation
\begin{equation*}
  \begin{gathered}
    \includegraphics[scale=1.0]{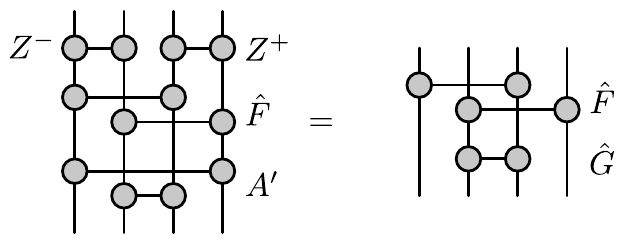}
  \end{gathered},
\end{equation*}
where we introduced the matrix
\begin{equation*}
  \hat{G} := \hat{F} \dunity_2  = \frac{1}{\sqrt{2}}
  \begin{pmatrix}
    \phantom{-}1 & 1 \\
    -1 & 1
  \end{pmatrix}
  .
\end{equation*}
Replacing the operations in the dashed box in the right diagram of \cref{fig:dst-3-orthogonal} by
the right hand side of \cref{fig:dst-3-orthogonal-detail} (in the appropriate size), we obtain a
recursion relation for the {$\dstiii$} that only contains orthogonal operations.

This completes all steps that are required to obtain a completely orthogonal recursion relation for
the $\dsti$ of size $n=2m-1=2^{k+1}-1$.  Let us summarize our final results: For the {$\dsti_n$}, we
arrive at a binary recursion, which can be represented diagrammatically, e.g.\ for ${n=7}$, as
\begin{subequations} \label{eq:dst-1o-alg}
  \begin{equation}
    \dstio_7 = 
    \begin{gathered}
      \includegraphics[scale=1.0]{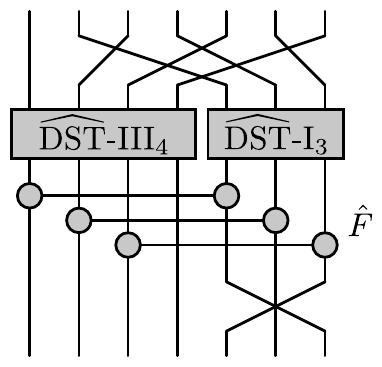}
    \end{gathered}.
  \end{equation}
  For arbitrary odd size $n=2m-1$, this can also be expressed as
  \begin{multline}
    \dstio_{2m-1} = \\
    \bar{L}_{2m-1} \big( \dstiiio_m \oplus \dstio_{m-1} \big) \hat{M}_{2m-1}.
  \end{multline}
\end{subequations}
Here, we defined the matrix
\[
  \hat{M}_{2m-1} = \frac{1}{\sqrt{2}}
  \begin{pmatrix}
    \unity_{m-1} & & \phantom{-}\dunity_{m-1} \\
     & \sqrt{2} & \\
    \unity_{m-1} & & -\dunity_{m-1} 
  \end{pmatrix},
\]
which is just the orthogonalized version of the matrix $B_{2m-1}$ from the non-orthogonal relation
\eqref{eq:dst-1-alg}.

For the {$\dstiii_n$}, we found the diagram
\begin{subequations} \label{eq:dst-3o-alg}
  \begin{equation}
    \dstiiio_8 = 
    \begin{gathered}
      \includegraphics[scale=1.0]{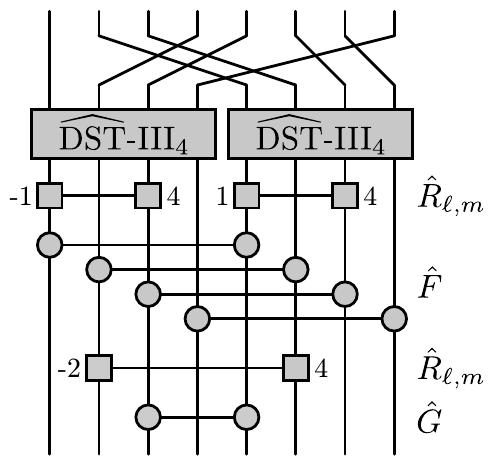}
    \end{gathered},
  \end{equation}
  which is given here as an example for $n=8$. Again this has a general expression for $n=2m$, given
  by
  \begin{multline}
    \dstiiio_{2m} = K_{2m} \big( \dstiiio_m \oplus \dstiiio_m \big) \\
    \times \big( \hat{Q}_m^- \oplus \hat{Q}_m^+ \big) \big( \dstiiio_2 \otimes \, \unity_m \big)
    \hat{N}_{2m}.
  \end{multline}
\end{subequations}
The newly introduced matrix $\hat{Q}_m^\pm$ is acting on pairs of components $(\ell-1,m-1-\ell)$,
$\ell=1,\dots,m/2-1$ via $\hat{R}_{\pm\ell,m}$ and therefore has a similar structure as $X_m^\pm$
defined in \cref{eq:x-matrix}.  The other new matrix $\hat{N}_{2m}$ couples the pair of components
$(m/2-1,3m/2-1)$ by $\hat{R}_{-m,2m}$ and the pairs of components $(m/2-1+\ell,3m/2-1-\ell)$,
$\ell=1,\dots,m/2-1$ via $\hat{G}$.

\begin{figure*}
  \centering
  \includegraphics[scale=1.0]{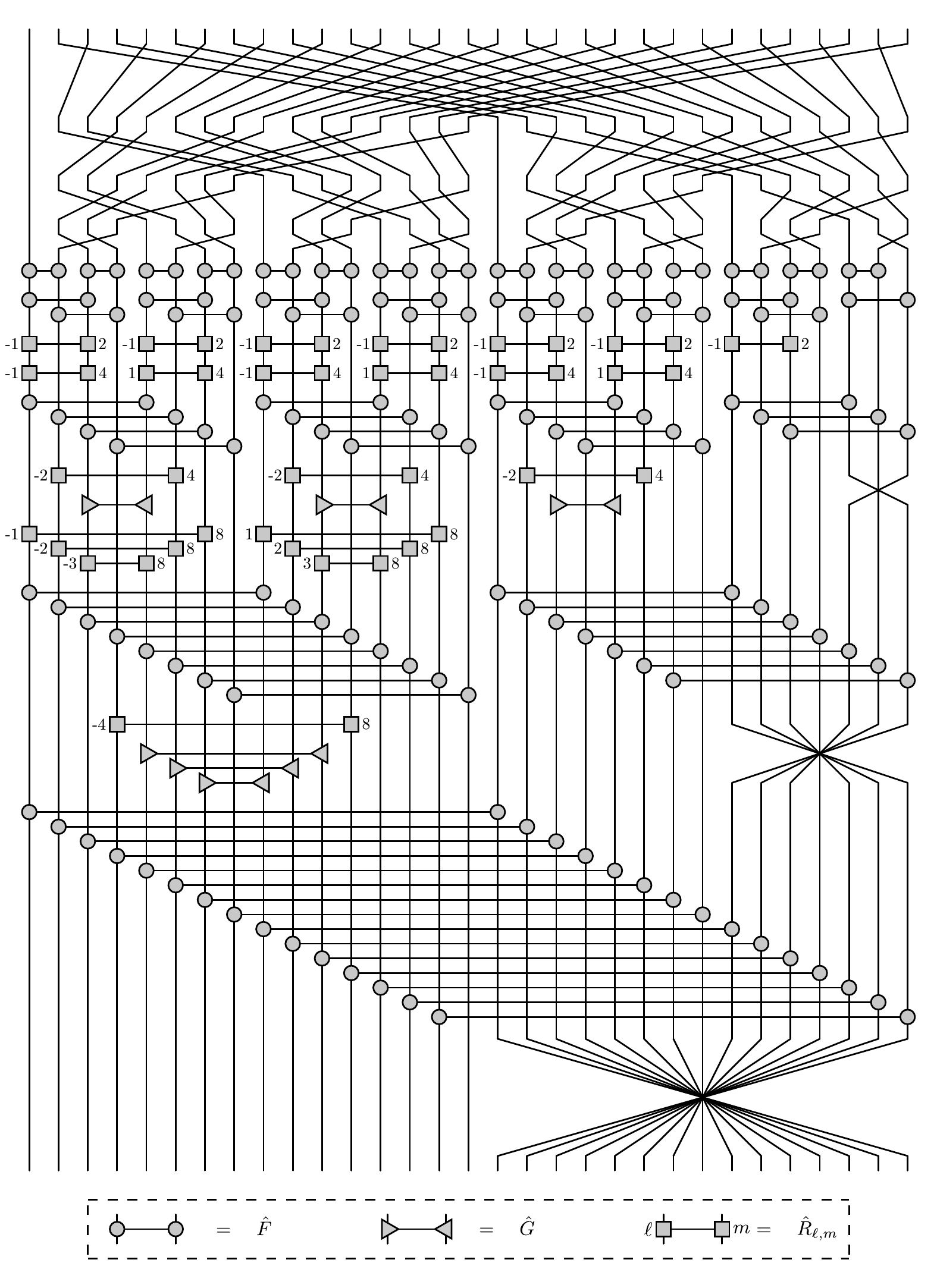}
  \caption{Diagrammatic representation of the $\dstio_{31}$ made up entirely from orthogonal
    operations acting on only two components. The diagram is obtained from the recursion relations
    \eqref{eq:dst-1o-alg} and \eqref{eq:dst-3o-alg}, together with the corresponding closing
    conditions \eqref{eq:closing-conditions}.}
  \label{fig:dst-1-complete}
\end{figure*}

The corresponding closing conditions for \cref{eq:dst-1o-alg,eq:dst-3o-alg} are
\begin{equation} \label{eq:closing-conditions}
  \dsti_1 = 1 \qtext{and} \dstiiio_2 = \hat{F},
\end{equation}
such that the complete recursion leads to a well-defined network of operations.  As an example, we
drew the complete network for the $\smash{\dstio_{31}}$ in \cref{fig:dst-1-complete}.

To calculate the computational complexity of the derived algorithm, denote by $\ci_n$ and $\ciii_n$
the number of elementary operations in the $\dstio_n$ and $\dstiiio_n$, respectively, neglecting
permutations.  From \cref{eq:dst-1o-alg}, we find
\[
  \ci_{2^{k+1}-1} = \ci_{2^k-1} + \ciii_{2^k} + (2^k - 1),
\]
where the last summand is the number of $\hat F$ matrices in each recursion step.  On the other hand,
\cref{eq:dst-3o-alg} implies
\[
  \ciii_{2^{k+1}} = 2 \ciii_{2^k} + (2^k-1) + 2^k + (2^k-1),
\]
the additional summands being the number of $\hat R_{\ell,m}$, $\hat F$, and $\hat G$ matrices per
recursion step, in that order.  Evaluating these formulae, anchored at $\ci_1=0$ and $\ciii_2=1$, we
obtain
\begin{subequations} \label{eq:complexity}
  \begin{gather}
    \ci_n = \frac 54 n \log_2 (n+1) - \frac {13}4 n + \frac 94 \log_2 (n+1) - \frac 14 \\
    \text{and} \quad \ciii_n = \frac 54 n \log_2 n - \frac 74 n + 2.
  \end{gather}
\end{subequations}

\section{Second quantization of diagrams}
\label{sec:diagram-quantization}
We will now outline a general method for performing second quantization of diagrammatic algorithms.
We shall discuss fermions only -- all results, however, extend naturally to the bosonic case
\cite{Ferris:14,Derezinski:13}.

Let us start with a basis transformation $U$ on the single-particle Hilbert space.  Its matrix
representation is given by
\begin{equation} \label{eq:single-particle-matrix}
  U \ket{a} = \sum_b \ket{b} \braopket{b}{U}{a} =: \sum_b U^{ba} \ket{b}.
\end{equation}
We can extend $U$ to a basis transformation $\Gamma_{\!U}$ of the multi-particle Hilbert space, by
having it leave the vacuum $\ket{\Omega}$ invariant and act linearly on creation operators
$c^\dagger$ \cite{Bogoliubov:58,Derezinski:13}.  This means that in the occupation number basis
\[
  \ket{k} = (c^\dagger_0)^{k_0} \!\cdots (c^\dagger_{n-1})^{k_{n-1}} \ket{\Omega},
  \quad k_a \in \braces{0,1},
\]
we define the \emph{second quantization} $\Gamma_{\!U}$ of $U$ by
\begin{equation} \label{eq:multi-particle-transformation}
  \Gamma_{\!U} \ket{k} :=
  \big(c^\dagger_{U\ket{0}}\big)^{k_0} \!\cdots \big(c^\dagger_{U\ket{n-1}}\big)^{k_{n-1}}
  \ket{\Omega},
\end{equation}
where the transformed mode
\begin{equation} \label{eq:mode-transformation}
  c^\dagger_{U\ket{a}} := \sum_b U^{ba} c^\dagger_b 
\end{equation}
creates a fermion in the state $U\ket{a}$.

\begin{figure*}
  \centering
  \includegraphics[scale=1.0]{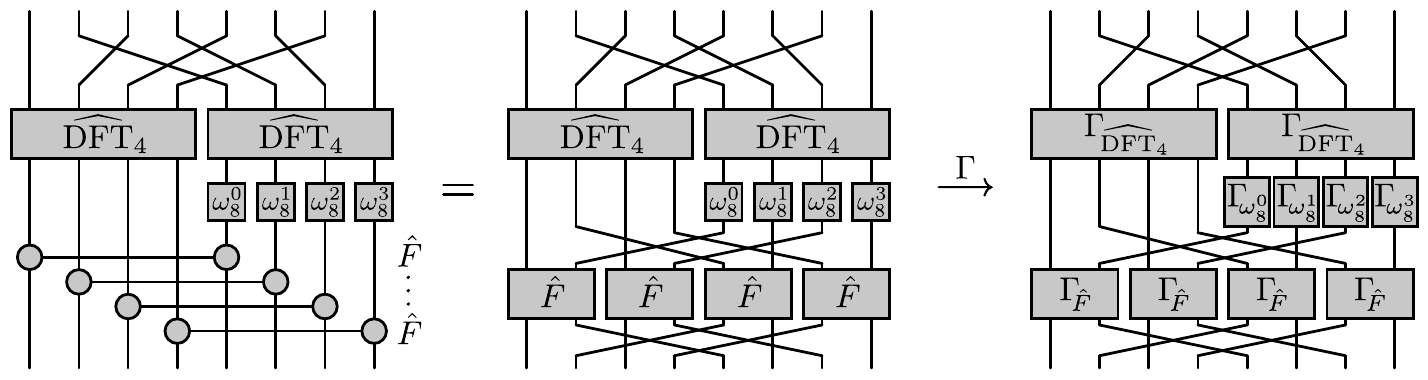}
  \caption{\label{fig:fft-2nd-quantization}Second quantization of the FFT diagram from
    \cref{eq:fft-diagram}.  First, we unravel the shorthand notation \eqref{eq:shorthand-hline}.  Then
    each block is replaced by its second quantization via
    \cref{eq:scalar-extension,eq:matrix-extension}.  The overall structure of the diagrams is
    unchanged because of the functorial relations \cref{eq:functoriality}, however, whenever two
    lines cross, we have to insert the fermion swap gate $\Gamma_{\!S}$, obtained in
    \cref{eq:fermion-swap}.}
\end{figure*}

Similarly to \cref{eq:single-particle-matrix}, we then have 
\[
  \Gamma_{\!U} \ket{k} = \sum_l \Gamma_U^{lk} \ket{l},
\]
where \cref{eq:multi-particle-transformation,eq:mode-transformation} can be used to obtain
\begin{multline} \label{eq:full-vacuum-expectation-value}
  \Gamma_{\!U}^{lk} = \Big\langle \Omega \Big| c^{l_{n-1}}_{n-1} \cdots c^{l_0}_0
  \Big(\sum_{b_0} U^{b_0 0} c^\dagger_{b_0} \Big)^{\!k_0} \cdots \\
  \times  \Big(\!\sum_{b_{n-1}} U^{b_{n-1} n-1} c^\dagger_{b_{n-1}} \Big)^{\!k_{n-1}}
  \Big| \Omega \Big\rangle.
\end{multline}
Denote now by
\begin{align*}
  L &= (L_0,\ldots, L_{p_l-1}) := \braces{a | l_a = 1} \qtext{and} \\
  K &= (K_0,\ldots, K_{p_k-1}) := \braces{a | k_a = 1}
\end{align*}
the (ordered) lists of occupied modes in $\ket{l}$ and $\ket{k}$, respectively, where $p_l$ and
$p_k$ are their total numbers.  Obviously, we have $\Gamma_{\!U}^{lk} = 0$ for $p_l \neq p_k$, since
then the modes in \cref{eq:full-vacuum-expectation-value} do not match up in pairs.  Let us thus
consider the case where $p_k = p_l =: p$.  Here, we have
\begin{multline*}
  \Gamma_{\!U}^{lk} = \!\!\!\!\!\!\sum_{\smash{b_0,\ldots,b_{p-1}}}\!\!\!\!\!
  U^{b_0 K_0} \!\cdots U^{b_{p-1} K_{p-1}} \\
  \times \braopket{\!\Omega}{c^{}_{L_{p-1}} \!\!\!\!\!\cdots c^{}_{L_0} c^\dagger_{b_0} \!\!\cdots
    c^{}_{b_{p-1}}}{\Omega\!},
\end{multline*}
where the expectation value on the right hand side again vanishes if $(b_0,\ldots,b_{p-1}) \neq L$ as
sets.  Since the summand also changes sign under odd permutations of $(b_0,\ldots,b_{p-1})$, we obtain
\begin{equation} \label{eq:multi-particle-matrix}
  \begin{split}
    \Gamma_{\!U}^{lk} &= \sum_{\pi \in S_p} \sgn(\pi) U^{L_{\pi(0)} K_0} \!\cdots U^{L_{\pi(p-1)} K_{p-1}} \\
    &= \det \parens{(U^{L_b K_a})_{0 \leq a,b < p}},
  \end{split}
\end{equation}
which shall serve as a recipe for the calculation of $\Gamma_{\!U}$.

The Slater determinant expression \eqref{eq:multi-particle-matrix} enables us to check the
functorial relations \cite{Derezinski:13}
\begin{equation} \label{eq:functoriality}
  \Gamma_{\!UU'} = \Gamma_{\!U}\Gamma_{\!U'} \qtext{and}
  \Gamma_{\!U \oplus U'} = \Gamma_{\!U} \otimes \Gamma_{\!U'}.
\end{equation}
Also, we see that second quantization preserves unitarity and orthogonality, since
\[
  \Gamma_{\!U^\dagger} = \Gamma_{\!U}^\dagger, \quad \Gamma_{\!U^{\mathrm T}} =
  \Gamma_{\!U}^{\mathrm T}, \qtext{and} \Gamma_{\!\unity_n} = \unity_{2^n}.
\]
We can now use \cref{eq:functoriality} to second-quantize the diagrams from the preceding
sections.  This amounts to replacing all single particle matrices $U$ by their respective second
quantizations $\Gamma_U$.  Each vertical line then represents an occupation number, hence the
vertical composition of blocks as on the left hand side of \cref{eq:single-particle-diagram-rules}
turns into a tensor contraction
\[
  \sum_{k_3, k_4} \Gamma_{\!B}^{k_1 k_2, k_3 k_4} \Gamma_{\!A}^{k_3 k_4, k_5 k_6}
  = (\Gamma_{\!B} \Gamma_{\!A})^{k_1 k_2, k_5 k_6},
\]
while the right hand side turns into the tensor product
\[
  \Gamma_{\!C}^{k_1 k_2, k_3 k_4} \Gamma_{\!D}^{k_5 k_6, k_7 k_8}
  = (\Gamma_{\!C} \otimes \Gamma_{\!D})^{k_1 k_2 k_5 k_6, k_3 k_4 k_7 k_8}.
\]
Note that this does not affect the \emph{structure} of the diagrams at all, but just amounts to a
reinterpretation of what they are supposed to mean.

Since only scalars $\alpha$ and $2 \times 2$-matrices $U$ appear in the diagrams we use, we can
explicitly evaluate \cref{eq:multi-particle-matrix} to obtain
\begin{gather}
  \Gamma_{\!\alpha} =
  \begin{pmatrix}
    1 &0 \\
    0 &\alpha
  \end{pmatrix} \qtext{and} \label{eq:scalar-extension} \\
  \Gamma_{\!U} =
  \begin{pmatrix}
    1 &      &     &\\
     &U^{11} &U^{10} &\\
     &U^{01} &U^{00} &\\
     & & &\det U
  \end{pmatrix}. \label{eq:matrix-extension}
\end{gather}
in the Kronecker basis $\braces{\ket{00},\ket{01},\ket{10},\ket{11}}$.

Finally, we have to give a meaning to the crossings of vertical lines, as in
\cref{eq:shorthand-hline}.  For a single particle, these are represented by the swap matrix
\[
  S := \begin{pmatrix} 0 &1 \\ 1 &0 \end{pmatrix},
\]
hence in the multi-particle case, we can apply \cref{eq:matrix-extension} to find
\begin{equation} \label{eq:fermion-swap}
  \Gamma_{\!S} = \begin{pmatrix} 1&&&\\ &0&1&\\ &1&0&\\
    &&&-1 \end{pmatrix},
\end{equation}
which correctly picks up a negative sign if two fermions switch places.

As an example, consider the normalized version of the FFT diagram in \cref{eq:fft-diagram}.  We can
recursively break it down, so that we only need the second quantization of $\hat F$ from
\cref{eq:def-f-hat}, given by the matrix
\[
  \Gamma_{\!\!\hat{F}} =
  \begin{pmatrix}
    1 &      &    &\\
     &- 1/\sqrt{2} &1/\sqrt{2} &\\
     &1/\sqrt{2} &1/\sqrt{2} &\\
     & & &-1
  \end{pmatrix}
\]
and local terms as in \cref{eq:scalar-extension}.  The entire process can be found in
\cref{fig:fft-2nd-quantization}, resulting in a diagram that exactly reproduces the \emph{spectral
  tensor network} by \cite{Ferris:14,Verstraete:09}.  Correspondingly, we can use the same rules on
the unitary decompositions \eqref{eq:dst-1o-alg} and \eqref{eq:dst-3o-alg} derived in the preceding
section, to obtain a quantum circuit implementing the {$\dsti$} for fermions and thus generalizing the
spectral tensor network for open boundary conditions.  Therefore, \cref{eq:complexity} gives an upper
bound for the \emph{quantum computational complexity} \cite{Nielsen:00} of the {$\dsti$} and {$\dstiii$}
of fermionic modes.  Note, however, that we omitted any permutations in the calculation leading to
\cref{eq:complexity}.  While this is fine if we deal with just a single particle, exchange statistics
need to be incorporated for multiple fermions, hence permutations need to be decomposed into gates
of type \eqref{eq:fermion-swap}.  This leads to additional $\sim \frac 76 n^2$ gates, where the
coefficient arises from the most naive decomposition and can likely be dropped to a smaller value.

\section{Discussion}
The present work is based on previous work by P\"uschel and Moura, who gave a purely algebraic
framework for the study of spectral transformations for various types of boundary conditions
\cite{Moura:08a,Moura:08b} and their recursive decomposition into sparse matrices
\cite{Moura:03,Moura:08c}.  Although this work is remarkable and very general, its practical use for
quantum mechanics is limited in so far as the building blocks of the resulting decomposition are not
unitary -- a crucial property of any transformation which is to be interpreted as a change of
orthonormal bases in a Hilbert space.  Since unitary variants of these Fourier transforms exist, it
is of course near at hand to expect that also the corresponding recursion relations can be
reformulated in terms of unitary building blocks, but carrying out such a unitarization could be
quite cumbersome.

In this paper, we have used a diagrammatic language to explicitly demonstrate said unitarization at the
example of a discrete sine transformation.  This led to a decomposition where the sparse matrices
are direct sums of unitary elementary operations, hence well suited for parallelization.  We expect
that other generalized Fourier transformations can be made unitary in a similar way, although then
the technicalities are probably even more involved.

The fact that the resulting recursion relations consist of block-diagonal unitaries is particularly
important when turning to many particles in the context of second quantization.  We have shown that
such a second-quantized version of a sine transformation for fermions can be obtained by replacing
the unitary building blocks of the diagram by appropriate second-quantized counterparts and suitable
modifications for line crossings, to implement the particle statistics \cite{Orus:14b}.  Doing so, we
arrive at a tensor network, whose structure is essentially the same as that of the original diagram,
a circumstance that has already been noted in the context of the ordinary DFT on a circle by Ferris
\cite{Ferris:14}.

Another great advantage of unitary building blocks becomes apparent when calculating local
expectation values with the resulting tensor network: as in the case of the MERA, a causal structure
emerges \cite{Vidal:08} and parts of the network that are not causally connected to the considered
region can be contracted to trivial operations.  Therefore, the effective complexity of the
computation of one- and two-point functions drops even further, scaling just linearly with the
system size.

The network thus makes it possible to numerically study boundary effects in one dimensional free
fermion models, which are exactly solvable by means of a spectral transformation. The Jordan-Wigner
transformation extends the applicability to spin-$\frac{1}{2}$ models, such as the XY model
\cite{Lieb:61}. Imposing the variational method on the tensor coefficients, while fixing the overall
topology, the proposed network could also provide a starting point for the simulation of a wider
class of models with non-cyclic boundary conditions. Furthermore, since Ferris' similar construction
for the DFT naturally generalizes to higher dimensions, we expect that this also holds for the
$\dsti$.

Finally, the observation that second quantization preserves the structure of diagrams, which can be
seen as some kind of \emph{Bohr correspondence principle} on a higher level, seems to be very
fundamental and is linked to the underlying category theory, as we shall discuss in a forthcoming
paper.

\bibliographystyle{apsrev4-1}
\bibliography{references.bib}

\end{document}